\documentclass[a4paper,pra,reprint,twocolumn,superscriptaddress]{revtex4}

\usepackage{ulem}
\usepackage{amssymb}
\usepackage{amsmath}
\usepackage{epsfig}
\usepackage{color}
\usepackage{graphics, graphicx}
\usepackage{bbold}
\usepackage{psfrag}
\usepackage{mathcomp}
\usepackage{amsmath}
\usepackage{amssymb}
\usepackage{mathrsfs}
\usepackage{subfigure}
\usepackage{verbatim}
\usepackage{xcolor,bm}
\usepackage[colorlinks,citecolor=blue,urlcolor=blue]{hyperref}

\usepackage{mathrsfs}

\begin{document}

\title{Synthetic topology and Floquet dynamic quantum phase transition in a periodically driven Raman lattice}
\author{De-Huan Cai}
\affiliation{CAS Key Laboratory of Quantum Information, University of Science and Technology of China, Hefei 230026, China}
\affiliation{CAS Center For Excellence in Quantum Information and Quantum Physics, Hefei 230026, China}
\affiliation{Bengbu University, Bengbu 233030, China}
\author{Wei Yi}
\email{wyiz@ustc.edu.cn}
\affiliation{CAS Key Laboratory of Quantum Information, University of Science and Technology of China, Hefei 230026, China}
\affiliation{CAS Center For Excellence in Quantum Information and Quantum Physics, Hefei 230026, China}

\begin{abstract}
Stimulated by the recent progress in engineering topological band structures in cold atomic gases, we study the dynamic topological phenomena for atoms loaded in a periodically driven optical lattice. When the frequency of the periodic modulation is low, the time-dependent Hamiltonian can be mapped to a two-dimensional topological insulator, with the discretized frequency components playing the role of an additional, synthetic dimension. In the high-frequency limit, we derive the effective Floquet Hamiltonian of the system, and reveal the occurrence of Floquet dynamic quantum phase transitions---an emergent topological phenomenon in the micromotion of the Floquet dynamics. Addressing the relation between the topology of the effective Floquet Hamiltonian and the presence of dynamic topological phenomena, we demonstrate that the topologically non-trivial nature of the Floquet Hamiltonian is a sufficient but not necessary condition for the onset of the Floquet dynamic quantum phase transition. We further discuss the relation of the topology of the Floquet Hamiltonian with the existence of dynamic skyrmion structures in the emergent momentum-time manifold of the micromotion, as well as the fate of these dynamic topological phenomena when the modulation frequency decreases away from the high-frequency limit. Finally, making use of the rich level structures of $^{171}$Yb atoms, we show that the system under study can be implemented in a one-dimensional Raman lattice where states in the $^1S_0$ ground-state manifold are coupled by Raman beams with periodically modulated amplitudes.
\end{abstract}

\maketitle

\section{Introduction}
Topological matter in solid-state materials exhibits robust properties that are protected by the global geometry of ground-state wave functions at equilibrium~\cite{MHasan,qixiaoliang}. With the fast progress of quantum simulators, many of these features have been observed in synthetic topological systems such as cold atoms~\cite{Bloch,ETHcoldatom14,Ketterle,Weitenberg2016,Greiner,yanbo}, photonics~\cite{KB+12,Rechtsman13,Cardano2017,XYXu2018,BWang2018,CChen2018}, solid spins~\cite{Lukin14,FRaman,bingsolid}, and superconducting qubits~\cite{qubitchiral,qubitqw,qubitDQPT}. The flexible controls available to the synthetic platforms further extend the scope of the study, enabling the identification of topological phenomena in non-equilibrium quantum dynamics~\cite{Rechtsman13,CSPRA,Lindner11,Levin13,DQPT1,zhaizheng,zhouqi,Doradynamic,Heyldynamic,linking1,ionDQPT,2ddqptatoms,slager2,slager1,DQPT3,schen,bandinver1,bandinver2,bandinver3,skyrmions1,skyrmions2,linking2,Photonic2,jad1,jad2,Photonic3,ftpcoldatoms,FDQPT1,FDQPT2,FDQPT3,FDQPT4,dqptqw,zhouFDQPT,kunkun}.
Examples of the dynamic topological phenomena include Hopf links~\cite{linking1,linking2} and dynamic skyrmions in the synthetic momentum-time manifold~\cite{skyrmions1,skyrmions2,schen}, band-inversion surfaces~\cite{bandinver1,bandinver2,bandinver3}, and dynamic quantum phase transitions (DQPT) characterized by dynamic topological order parameters~\cite{dqptqw,zhouFDQPT,Heyldynamic,DQPT3}.
These dynamic topological constructions are closely related to the topology of Hamiltonians driving the dynamics, and often offer valuable means for the detection of topological invariants of the relevant static models. For instance, in a typical quantum quench, the system is initialized in the ground state of a topologically trivial Hamiltonian, and subsequently evolves under a final Hamiltonian. The topology of the final Hamiltonian can be subsequently deduced from the presence (or the lack thereof) and the pattern of dynamic topological structures.

Alternatively, dynamic topological phenomena also emerge in periodically driven Floquet systems~\cite{Rechtsman13,dqptqw,zhouFDQPT,CSPRA,Lindner11,Levin13,zhaizheng,zhouqi,Photonic2,Photonic3,ftpcoldatoms,FDQPT1,FDQPT2,FDQPT3,FDQPT4,DQPT3,slager1}. In the simplest Floquet configuration, a Floquet operator $U$ is repeatedly acted on the instantaneous state of the system, such that the overall time evolution is seen as a stroboscopic simulation of the continuous-time evolution driven by an effective Hamiltonian $H_{\text{eff}}$, with $U=e^{-iH_{\text{eff}}T}$~\cite{EckardtFloquet1,EckardtFloquet2,EckardtFloquet3}. Here $T$ is the duration of a single discrete time step. Since $H_{\text{eff}}$ can possess band topology, the Floquet dynamics can be regarded as a discrete-time quench process of a topological Hamiltonian. Along this vein of thought, dynamic skyrmions and DQPTs have recently been observed in discrete-time quantum walks of single-photons and cold atoms~\cite{dqptqw,FDQPT4,kunkun}.
Here DQPT is the temporal analog of continuous phase transitions, characterized by non-analyticities in the time evolution of physical observables. Formally, the DQPT occurs as the Loschmidt amplitude $G(t)=\langle \psi(0)|\psi(t)\rangle$ vanishes at critical times of the evolution, where $|\psi(t)\rangle$ is the time-evolved state.

Due to the generality of its definition, DQPTs also exist in Floquet dynamics with time-periodic Hamiltonians~\cite{zhouFDQPT,FDQPT1,FDQPT2}, without invoking the concept of quench dynamics.
Here a general Floquet operator is written as $U=\mathcal{T} e^{-i\int_0^T dt H(t)}$, where $\mathcal{T}$ is the time-ordering operator and $H(t)$ is a general time-dependent Hamiltonian. While it has been shown that a Floquet DQPT can occur in the micromotion within each Floquet period~\cite{zhouFDQPT,FDQPT1,FDQPT2}, the relation of Floquet DQPTs with other dynamic topological phenomena has seen only limited study, and its experimental observation has so far been limited to a parametric demonstration using zero-dimensional solid spins~\cite{zhouFDQPT}.

In this work, we study emergent topological structures in a periodically driven, one-dimensional lattice model, which derives from the recently realized Raman lattice potentials~\cite{2Dsoc,Gyubongjo18,xjliu1D,xjliu2D,jspan}.
When the driving frequency is low, the discretized frequency modes of the Fourier-transformed Hamiltonian takes the form of a synthetic dimension~\cite{Platero,bgliu,dhcai}, and the resulting two-dimensional Hamiltonian can be mapped to that of a topological insulator, exhibiting quantized transport in the time domain.
We then focus on the high-frequency regime, where the effective Floquet Hamiltonian can be derived in a perturbative fashion. We show that Floquet DQPTs emerge in the micromotion of the Floquet dynamics, and discuss their close relations with the topology of the Floquet operator (or, equivalently, that of $H_{\text{eff}}$), as well as with other forms of dynamic topological phenomena such as the dynamic skyrmions.
As the modulation frequency decreases from the high-frequency limit, however, the simple picture above no longer holds, due to complications in the micromotion.
Nevertheless, the DQPT can still occur away from the high-frequency regime.
This is consistent with previous studies of Floquet DQPTs in systems where the effective Hamiltonians can be analytically derived, and are therefore also beyond the higher-frequency limit~\cite{zhouFDQPT,FDQPT1,FDQPT2,FDQPT3}.
With the mature controls amenable to alkaline-earth(-like) atoms following decades of efforts in developing optical lattice clocks, our proposed configuration is readily implementable under current experimental conditions. As an example, we discuss in detail how the time-dependent Raman potentials  can be implemented using $^{171}$Yb atoms.


Our work is organized as the following. In Sec. II, we write down the model Hamiltonian featuring time-periodic Raman potential, and derive the corresponding tight-binding Hamiltonian. We then study the two-dimensional topological model in the synthetic frequency-momentum domain in the low-frequency limit in Sec. III, followed by discussions of emergent topological phenomena in the high-frequency regime in Sec. IV. We discuss the relation between the Floquet DQPT,
the topology of the Floquet Hamiltonian, and the dynamic skyrmion structures in this section. In Sec. V, we present in detail how to implement the proposed configuration using $^{171}$Yb atoms. Finally, we conclude in Sec. VI.

\section{model}

We consider a two-component atomic gas in a one-dimensional(1D), periodically driven Raman lattice, with the Hamiltonian (we set $\hbar=1$ throughout the work)
\begin{align}
 H = & \frac{p^{2}_x}{2m} + V_{\text{latt}}(x) + \frac{\delta}{2}(|\uparrow\rangle\langle\uparrow|-|\downarrow\rangle\langle\downarrow|)\nonumber\\
 &+ \Big\{[M_1(x,t)+iM_2(x,t)]|\uparrow\rangle\langle\downarrow| + \text{H.c.}\Big\}. \label{eq:RamanH}
\end{align}
Here $\sigma\in\{|\uparrow\rangle,|\downarrow\rangle\}$ label the two spin species,
$p^{2}_x/2m$ is the kinetic term along the lattice direction $x$, the lattice potential $V_{\text{latt}}(x) = V_0 \cos^2 (\pi k_0 x)$, and the time-periodic Raman potentials $M_1(x,t) = M_1\sin(\omega t)\cos(\pi k_0 x)$ and $M_2(x,t) = M_2\cos(\omega t) e^{i\pi k_0 x}$, where $\omega$ is the modulation frequency and $\delta$ is an effective Zeeman field. We assume that the other two spatial dimensions are tightly confined and hence frozen.

Following Refs.~\cite{xjliu1D,xjliu2D,jspan}, it is straightforward to derive the single-band, tight-binding Hamiltonian, now with a time-periodic Raman potential
\begin{align}
H_{\text{T}} = & -t_s\sum_{j}(c^{\dag}_{j,\uparrow} c_{j\pm 1,\uparrow} - c^{\dag}_{j,\downarrow} c_{j\pm 1,\downarrow}) + m_z \sum_{\sigma,j}\xi_{\sigma}c^{\dag}_{j,\sigma} c_{j,\sigma}\nonumber\\
& + [\sum_{j}t_{so}\sin(\omega t)(c^{\dag}_{j,\uparrow} c_{j-1,\downarrow} - c^{\dag}_{j,\uparrow} c_{j+1,\downarrow}) + \text{H.c.}]\nonumber\\
& + [\sum_{j} it'_{so}\cos(\omega t)(c^{\dag}_{j,\uparrow} c_{j-1,\downarrow} - c^{\dag}_{j,\uparrow} c_{j+1,\downarrow}) + \text{H.c.}]\nonumber\\
& - (\sum_{j} t''_{so}\cos(\omega t)c^{\dag}_{j,\uparrow} c_{j,\downarrow} + \text{H.c.} ). \label{eq:HTI}
\end{align}
Here $c_{j,\sigma}$ is the atomic annihilation operator of spin $\sigma$ on the $j$th site, $m_z = \frac{\delta}{2}$, $\xi_{\uparrow,\downarrow} = \pm 1$,
$t_s$ is the nearest-neighbor hopping rate, $t_{so}$ is the spin-orbit coupling associated with the Raman potential $M_1$, while $t'_{so}$ and $t''_{so}$ are those associated with $M_2$.
Such a periodically driven one-dimensional Raman lattice hosts rich dynamic topological phenomena. While our main focus of the work is in the high-frequency regime, for completeness, we first consider an adiabatically slow driving field, where the physics of interest is captured by the well-known topological Thouless pumping~\cite{thouless,pumpexp1,pumpexp2}.
We also note that, in the extreme high-frequency regime, the driving field necessarily couples higher bands, and the single-band model (\ref{eq:HTI}) no longer applies. We therefore require $V_0\gg\omega\gg t_s$ for our high-frequency regime, where the large band gap suppresses the high-band population.


\begin{figure}[tbp]
  \centering
  \includegraphics[width=9cm]{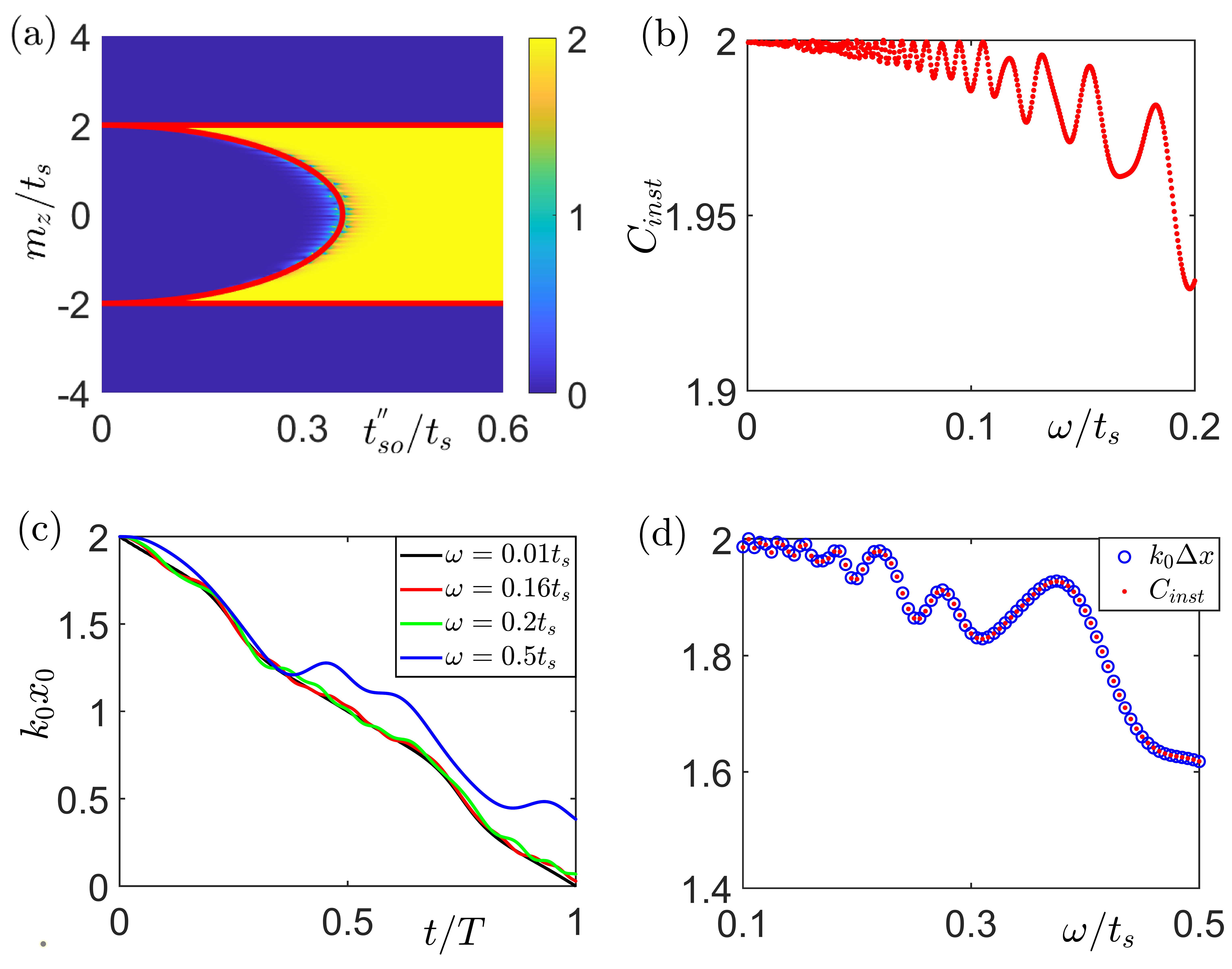}
  \caption{(a) Chern number as a function of $m_z$ and $t''_{so}$ in the low-frequency limit, where the red line indicates the phase boundary.
  (b) $C_{\text{inst}}$ as a function of $\omega$.
  With increasing $\omega$, $C_{\text{inst}}$ deviates from the quantized Chern number, indicating the breakdown of the adiabatic condition.
  (c) The trajectory of the Wannier center over one period of driving, in units of the lattice constant $1/k_0$. (d) The Wannier-center shift $\Delta x$ in one period, and $C_{\text{inst}}$ as functions of $\omega$. Their numerical results essentially overlap with each other. For calculations in (b), (c), and (d), we take $m_z=0$, $t''_{so}=1.0t_s$, $t_s\approx 0.12E_R$, and $t_{so}=t'_{so}=0.2t_s$. These in turn are determined by taking $V_0=4E_R$ and $M_1 = M_2 = 0.8E_R$ in the original Hamiltonian (\ref{eq:RamanH}).}
  \label{fig:fig1}
\end{figure}

\section{Synthetic topology in the low-frequency limit}

In this section, we review the description of the system in the low-frequency adiabatic limit, where the physics is well captured by the topological Thouless pumping.
This sets the stage for discussions of the Wannier-center shift beyond the low-frequency limit, and complements our study of the dynamic topological phenomena
in the high-frequency limit in Sec.~IV.

To gain insight into the time-periodic Hamiltonian (\ref{eq:HTI}), we perform a Fourier transform over time (see Appendix for more details), with the resulting Hamiltonian~\cite{Platero,bgliu,dhcai}
\begin{align}
H_{\text{FT}} = & m_z \sum_{n,j}(c^{\dag}_{n,j,\uparrow} c_{n,j,\uparrow} - c^{\dag}_{n,j,\downarrow} c_{n,j,\downarrow})\nonumber\\
& -\sum_{n,j}t_s(c^{\dag}_{n,j,\uparrow} c_{n,j\pm 1,\uparrow} - c^{\dag}_{n,j,\downarrow} c_{n,j\pm 1,\downarrow})\nonumber\\
& + [\sum_{n,j}\frac{i}{2}t_{so}(c^{\dag}_{n,j,\uparrow} c_{n-1,j-1,\downarrow} - c^{\dag}_{n,j,\uparrow} c_{n+1,j-1,\downarrow}\nonumber\\
& +c^{\dag}_{n,j,\uparrow} c_{n+1,j+1,\downarrow} - c^{\dag}_{n,j,\uparrow} c_{n-1,j+1,\downarrow})+ \text{H.c.}]\nonumber\\
& + [\sum_{n,j}\frac{i}{2}t'_{so}(c^{\dag}_{n,j,\uparrow} c_{n-1,j-1,\downarrow} + c^{\dag}_{n,j,\uparrow} c_{n+1,j-1,\downarrow}\nonumber\\
& -c^{\dag}_{n,j,\uparrow} c_{n+1,j+1,\downarrow} - c^{\dag}_{n,j,\uparrow} c_{n-1,j+1,\downarrow})+ \text{H.c.}]\nonumber\\
& - [\sum_{n,j}\frac{1}{2} t''_{so}(c^{\dag}_{n,j,\uparrow} c_{n+1,j,\downarrow} + c^{\dag}_{n,j,\uparrow} c_{n-1,j,\downarrow}) + \text{H.c.}]\nonumber\\
& + \sum_{n,j}n\omega(c^{\dag}_{n,j,\uparrow} c_{n,j,\uparrow} + c^{\dag}_{n,j,\downarrow} c_{n,j,\downarrow}), \label{eq:FloquetHTI}
\end{align}
where $n$ is the discrete frequency index (or, equivalently, the Floquet band index), and $c_{n,j,\sigma}=e^{in\omega t}c_{j,\sigma}$ is identified as the annihilation operator in the corresponding frequency mode $n$ (or the Floquet band $n$).

Equation (\ref{eq:FloquetHTI}) resembles a two-dimensional tight-binding model, if we regard the Floquet band index as an extra, synthetic dimension. The last term of (\ref{eq:FloquetHTI}) describes a potential, linear in the lattice site, along the synthetic dimension. It is similar to the potential an electron feels under a constant electric field of strength $\omega$~\cite{Platero,bgliu}. This tilting potential breaks the translational symmetry, but can be negligibly small in the low-frequency limit, with $\omega\ll t_s, t_{so}, t'_{so}, t''_{so}$.

Assuming such a low-frequency limit, let us first drop the term entirely. The lattice translational symmetry is then restored, such that we may write the Hamiltonian in the synthetic two-dimensional quasimomentum space
\begin{align}
H_{\text{FH}}=\sum_{k_t,k_x}\left(\begin{matrix} c^{\dag}_{k_t,k_x,\uparrow} & c^{\dag}_{k_t,k_x,\downarrow} \end{matrix}\right)H(k_t,k_x) \left(\begin{matrix} c_{k_t,k_x,\uparrow} \\ c_{k_t,k_x,\downarrow}\end{matrix}\right),
\end{align}
where $c_{k_t,k_x,\sigma} = \frac{1}{\sqrt{N_{t}N_{x}}}\sum_{k_t,k_x}e^{-ink_t}e^{-ijk_x}c_{n,j,\sigma}$, $(k_t,k_x) \in [-\pi,\pi)$ constitute an effective first Brillouin zone (EFBZ). Here the dimensionless quasimomentum $k_x$ is defined as the actual quasimomentum in units of $k_0$, and $N_t$ and $N_x$ denote the size of the synthetic frequency dimension and the spatial lattice dimension, respectively.
The Bloch Hamiltonian $H(k_t,k_x)$ is given by
\begin{align}
H(k_t,k_x) &= d_x\sigma_x + d_y\sigma_y + d_z\sigma_z \nonumber\\
&= \cos(k_t)[2t'_{so}\sin(k_x) - t''_{so}]\sigma_x + 2t_{so}\sin(k_t)\cdot \nonumber\\
&\sin(k_x)\sigma_y + [m_z - 2t_s\cos(k_x)]\sigma_z, \label{eq:Hktkx}
\end{align}
where $\sigma_{x,y,z}$ are Pauli matrices in the basis of the spins.

A central observation in this low-frequency limit is that $H(k_t,k_x)$ can have a nontrivial band topology, as it can be mapped to a topological insulator. The band topology is characterized by the Chern number
\begin{align}
C = \int_{\text{EFBZ}}(\bm{n}_d\times\partial_{k_t}\bm{n}_d)\cdot\partial_{k_x}\bm{n}_d dk_t dk_x, \label{eq:Chernnumber}
\end{align}
where $\bm{n}_d = \bm{d}/|\bm{d}|$ and $\bm{d}=(d_x,d_y,d_z)$.
In Fig.~\ref{fig:fig1}(a), we show the numerically evaluated Chern number as a function of $t''_{so}$ and $m_z$, where the region in yellow (blue) is topologically non-trivial (trivial). The topological phase boundary is given by $m_z = \pm 2 t_s$ or $(\frac{m_z}{2 t_s})^2 + (\frac{t''_{so}}{2t'_{so}})^2 = 1$ [red lines in Fig.~\ref{fig:fig1}(a)], where the band gap closes.

Under realistic conditions, the driving frequency $\omega$ is inevitably finite, it is therefore helpful to study situations where the low-frequency, adiabatic condition is not strictly met. This is reflected in the trajectory of the Wannier center, calculated from the time-evolved states under increasing modulation frequencies.

For the dynamics driven by Hamiltonian (\ref{eq:HTI}), we denote the time-evolved state in the momentum space as $|u^{\pm}_{k}(t)\rangle$, and replace $\bm{n}_d$ in Eq.~(\ref{eq:Chernnumber}) with $\bm{n}(t)$, where $\bm{n}(t)=\text{Tr}(\rho(t)\bm{\sigma})$, and $\rho(t)=|u^{-}_{k}(t)\rangle\langle u^{-}_{k}(t)|$ with $\bm{\sigma}=(\sigma_x,\sigma_y,\sigma_z)$.  This enables the calculation of a quantity, which we label as $C_{\text{inst}}$ and plot in Fig.~\ref{fig:fig1}(b).
Apparently, when the modulation frequency $\omega$ approaches zero, $C_{\text{inst}}$ approaches the actual quantized topological invariant in the adiabatic limit. Whereas oscillatory behavior is observed as $\omega$ increases, and $C_{\text{inst}}$ is in general not quantized.

Physically, $C_{\text{inst}}$ reflects the shift of the Wannier center after a single period of modulation. To see this, we introduce the position operator for atoms in the optical lattice with $N_x$ sites as~\cite{slager1,positionoperator,Wannier}
\begin{align}
X = \sum_{j,\sigma}c^{\dag}_{j,\sigma}|0\rangle e^{i\Delta_{k}(j+x_0)}\langle 0|c_{j,\sigma}, \label{eq:positionoperator}
\end{align}
where $x_0$ is the Wannier center, defined to be the atomic position relative to the site center, and $\Delta_{k}=2\pi/N_{x}$. The corresponding position operator in the momentum space is
$X = \sum_{k,\sigma}c^{\dag}_{k+\Delta_k,\sigma}|0\rangle\langle 0|c_{k,\sigma}$, where $c_{k,\sigma} = \frac{1}{\sqrt{N_{x}}}\sum_{j}e^{-i(j+x_0)}c_{j,\sigma}$.

Defining the projection operator of the lower band $P=\sum_{k}|u^-_{k}(t)\rangle\langle u^-_k(t)|$, we have the projected position operator
$X_{P} = PXP$.
Raising $X_{P}$  to its $N_x$-th power, we have
$(X_{P})^{N_x} = WP$, where the Wilson loop $W$ is defined as~\cite{Wannier}
\begin{align}
W = \langle u^{-}(2\pi)|u^{-}(2\pi-\Delta_k)\rangle...\langle u^{-}(\Delta_k)|u^{-}(0)\rangle.
\end{align}
Since $W$ is a unitary $e^{i2\pi\theta}$, the eigenvalues $\lambda_j$ of $X_P$ can be derived by evaluating the Wilson loop, with
$\lambda_j= e^{i\Delta_{k}(\theta+j)}$ ($j=1,\ldots, N_x$). Comparing with Eq.~(\ref{eq:positionoperator}), we immediately see that the Wannier center $x_0=\theta$.

In Fig.~\ref{fig:fig1}(c), we show the trajectory of the Wannier center over one period $T$ under different driving frequencies.
In the adiabatic limit, the Wannier center shifts over two lattice sites, consistent with the Chern number in the adiabatic limit [see Fig.~\ref{fig:fig1}(b) and black lines in Fig.~\ref{fig:fig1}(c)]. The situation is essentially the same as that of the Thouless pumping~\cite{thouless,pumpexp1,pumpexp2}.
With increasing $\omega$, the shift of Wannier center over one period is no longer quantized in lattice sites, as shown in Fig.~\ref{fig:fig1}(c), indicating the breakdown of the adiabatic condition. However, the overall shift of the Wannier center at $t=T$ is exactly $C_{\text{inst}}$ in Fig.~\ref{fig:fig1}(b) under the same parameters, see Fig.~\ref{fig:fig1}(d). Note that quantized Wannier-state pumping is reported in a previous study~\cite{wannierprl}. Therein, the transport is driven by a different mechanism (which persists beyond the low-frequency limit), and takes places in the momentum space instead of the real lattice space.

\begin{figure}[tbp]
  \centering
  \includegraphics[width=8.5cm]{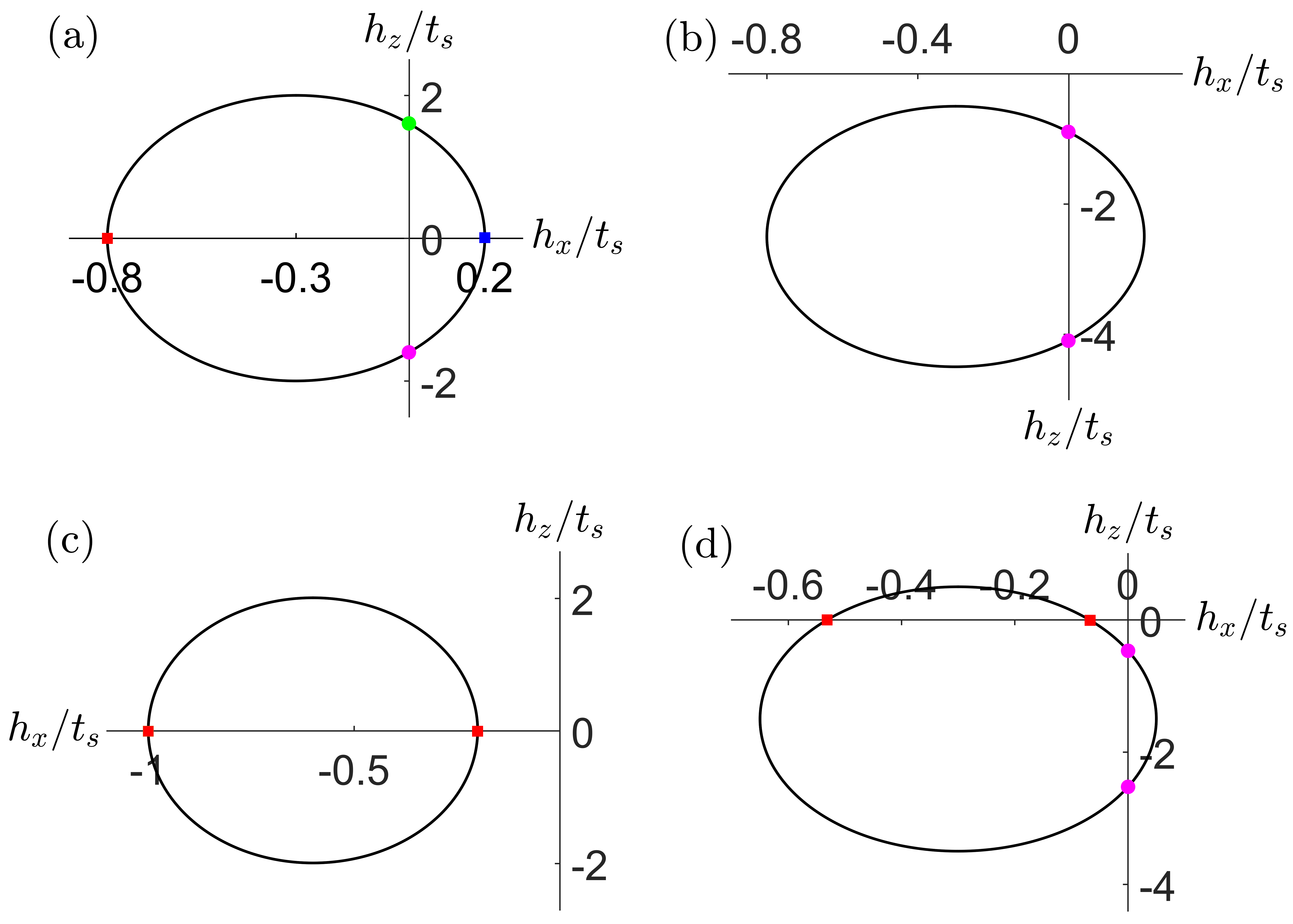}
  \caption{The elliptic trajectory of $\bm{h}$ as the dimensionless quasimomentum $k$ goes around the Brillouin zone. (a) Trajectory for $\nu=1$ for topologically nontrivial $H_{\text{eff}}$,
   with $m_z=10t_s$, $t_{so}=0.4t_s$, $t'_{so}=0.1t_s$, and $t''_{so}=0.6t_s$. (b) Trajectory with $\nu=0$. The parameters are the same as in (a) except for $m_z=7.5t_s$. (c) Trajectory with $\nu=0$. The parameters are $m_z=10t_s$, $t_{so}=t'_{so}=0.2t_s$, and $t''_{so}=1.2t_s$. (d) Trajectory with $\nu=0$. The parameters are $m_z=8.5t_s$, $t_{so}=0.25t_s$, $t'_{so}=0.1t_s$, and $t''_{so}=0.6t_s$. The magenta and green dots correspond to two different types of fixed points, and the red and blue square dots correspond to critical quasimomenta $k_c$ (see also Fig.~\ref{fig:fig3}). For our calculations, we take $V_0 = 8E_R$, so $t_s\approx 0.0375E_R$, and we take $\omega = 20t_s$. }
  \label{fig:fig2}
\end{figure}

%

\section{Dynamic topological phenomena in the high-frequency limit}

In the high-frequency limit, the system dynamics features emergent topological phenomena such as the Floquet DQPTs and dynamic skyrmions. To understand the condition of their occurrence, we first drive the effective Floquet Hamiltonian of the Bloch Hamiltonian in Eq.~(\ref{eq:BlochHkt}) using the high-frequency expansion.

\subsection{Effective Hamiltonian}

As the system has lattice translational symmetry along $x$, dynamics in different quasimomentum $k$ sectors are decoupled. Under the periodic modulation, the stationary states of the Floquet dynamics in each $k$ sector can be written as~\cite{zhouFDQPT,EckardtFloquet2}
\begin{align}
|\psi^{\pm}_{k}(t)\rangle = e^{-i\epsilon^{\pm}_{k} t}U^{k}_{F}(t)|\mu^{\pm}_{k}\rangle,\label{eq:F-dynamics}
\end{align}
where $\epsilon^{\pm}_{k}$ and $|\mu^{\pm}_k\rangle$ are the eigenenergies and eigenstates of the effective Floquet Hamiltonian $H_{\text{eff}}$, with $H_{\text{eff}}|\mu^{\pm}_k\rangle=\epsilon^{\pm}_{k}|\mu^{\pm}_k\rangle$ ($\pm$ are the band indices). Here $H_{\text{eff}}$ is defined as $U=e^{-iH_{\text{eff}}T}$, and $U$ is the time-evolution operator under $H(k,t)$ over each period $T=2\pi/\omega$. Formally therefore, $U=\mathcal{T}e^{-i\int_0^T H(k,t)dt}$.
According to the Floquet theorem, the micromotion operator $U^{k}_F(t)$ satisfies $U^{k}_{F}(t)=U^{k}_{F}(t+nT)$.

In the high-frequency regime with $2\omega \gg t''_{so}/2, |t'_{so}-t_{so}|, |m_z - \frac{\omega}{2}|$, we frist take a gauge transformation $U_{R}(t)=e^{i\omega (\sigma_0 -\sigma_z)t/2}$ on the time-dependent Bloch Hamiltonian (see Appendix for detail), and define the micromotion operator $U^{k}_{R,F}(t)=U^{\dag}_{R}(t)U^{k}_{F}(t)$. Here $\sigma_0$ is an identity $2\times 2$ matrix. Then, we can directly apply the high-frequency expansion to the first-order correction~\cite{EckardtFloquet2,EckardtFloquet3}
\begin{align}
H_{\text{eff}}& \cong H_0 + \sum_{m\neq 0}\frac{H_m H_{-m}}{m\omega},\label{eq:approHeff}\\
U^{k}_{R,F}(t)&\cong 1-\sum_{m\neq 0}\frac{e^{i m\omega t}}{m\omega}H_m.\label{eq:URFkt}
\end{align}
Explicit forms of $H_m$ are given in the Appendix.
It is then straightforward to derive
\begin{align}
H_{\text{eff}} \cong h_x(k)\sigma_x + h_z(k)\sigma_z + \frac{\omega}{2}\sigma_0,\label{eq:Heff}
\end{align}
and
\begin{align}
&U^{k}_{F}(t) \equiv U_{R}(t)U^{k}_{R,F}(t) \nonumber\\
&\cong\left[
   \begin{matrix}
     1 & -\frac{(t'_{so}-t_{so})\sin(k)-t''_{so}/2}{2\omega} e^{2i\omega t} \\
     \frac{(t'_{so}-t_{so})\sin(k)-t''_{so}/2}{2\omega} e^{-i\omega t} & e^{i\omega t} \\
   \end{matrix}
 \right]. \label{eq:micromotion}
\end{align}
Here $h_x (k)= (t'_{so}+t_{so})\sin(k)-t''_{so}/2$ and $h_z (k)= m_z - \omega/2 - 2t_s\cos(k) + [(t'_{so}-t_{so})\sin(k)-t''_{so}/2]^2/2\omega$.

From the effective Floquet Hamiltonian (\ref{eq:Heff}), we then derive the quasienergies and the eigenstates as
\begin{align}
\epsilon^{\pm}_{k}= \frac{\omega}{2} \pm \sqrt{h^{2}_{x}(k) + h^{2}_{z}(k)} = \frac{\omega}{2} \pm |\epsilon_k|,
\end{align}
and
\begin{align}
|\mu^{+}_{k}\rangle &= \left(\begin{matrix} \frac{h_{x}(k)}{\sqrt{h^{2}_{x}(k) + (|\epsilon_k|-h_{z}(k))^2}}\\ \frac{|\epsilon_k|-h_{z}(k)}{\sqrt{h^{2}_{x}(k) + (|\epsilon_k|-h_{z}(k))^2}}\end{matrix}\right),\\
|\mu^{-}_{k}\rangle &= \left(\begin{matrix}-\frac{|\epsilon_k|-h_{z}(k)}{\sqrt{h^{2}_{x}(k) + (|\epsilon_k|-h_{z}(k))^2}}\\ \frac{h_{x}(k)}{\sqrt{h^{2}_{x}(k) + (|\epsilon_k|-h_{z}(k))^2}}\end{matrix}\right).
\end{align}

Further, $H_{\text{eff}}$ has chiral symmetry, with $\sigma_{y}H_{\text{eff}}\sigma_{y}=-H_{\text{eff}}$. It follows that, as the quasimomentum $k$ runs through the Brillouin zone, the eigenvector $\bm{h}=(h_x,h_z)$ traces an elliptic trajectory on the $x$--$z$ plane. Importantly, the effective Floquet Hamiltonian is topological if the trajectory encloses the origin, as illustrated in Fig.~\ref{fig:fig2}(a). This is captured by the winding number
\begin{align}
\nu = \frac{1}{2\pi}\int^{\pi}_{-\pi}\bm{n}_h(k)\times\frac{d}{dk}\bm{n}_h(k) dk,
\end{align}
where $\bm{n}_h=\bm{h}/|\bm{h}|$. We also show in Figs.~\ref{fig:fig2}(b)-~\ref{fig:fig2}(d), some typical trajectories for $\nu=0$.
While the non-trivial band topology of $H_{\text{eff}}$ gives rise to topological edge modes under open boundary conditions, it also leads to emergent topological phenomena in Floquet dynamics under periodic boundary conditions.


\begin{figure}[tbp]
  \centering
  \includegraphics[width=8.5cm]{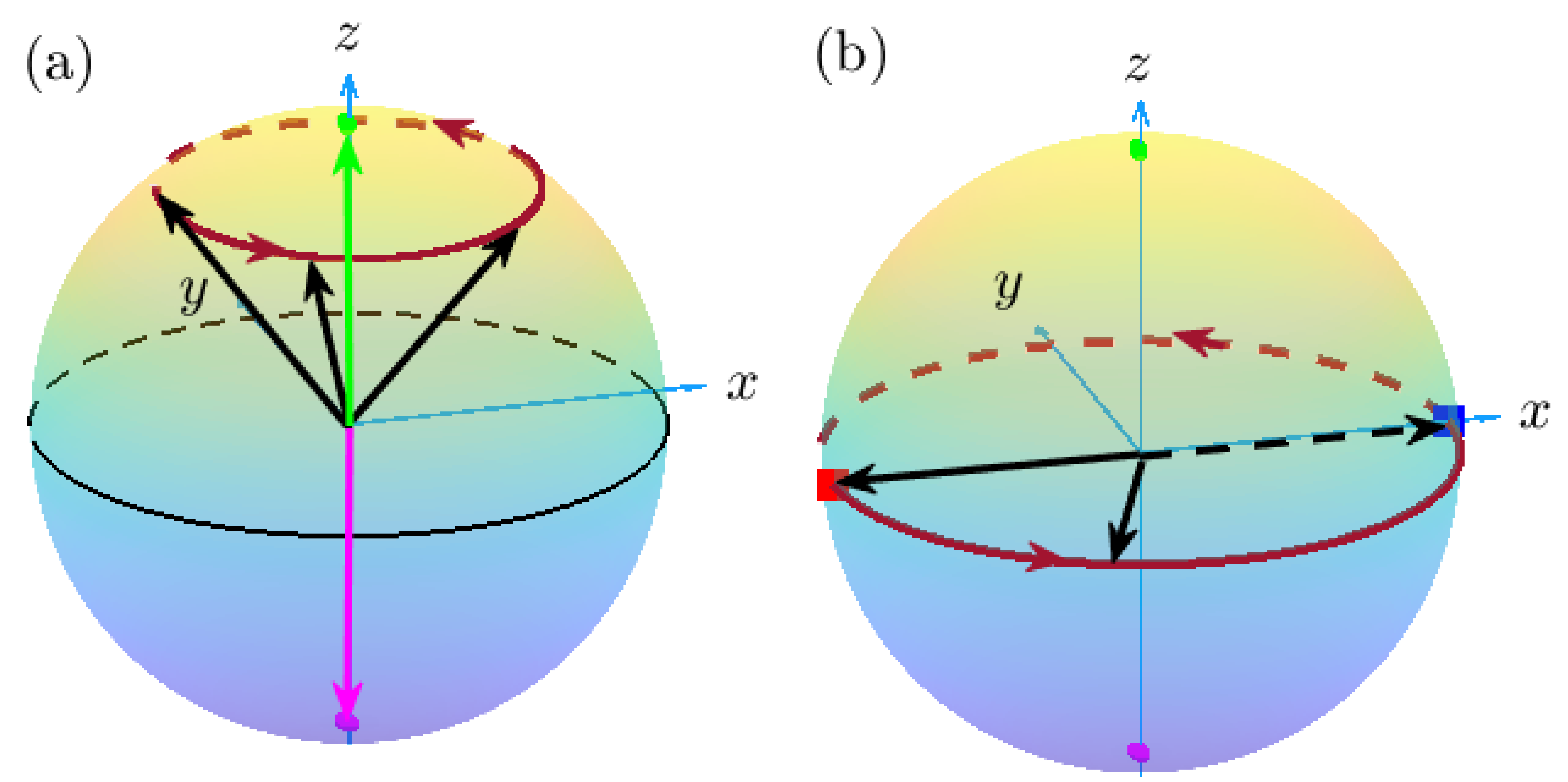}
  \caption{Schematic illustration of the spin dynamics $\bm{n}_{h}(k,t)$ on the Bloch sphere, in the high-frequency limit.
  (a) $\bm{n}_{h}(k,t)$ rotates around the $z$ axis in general. Two types of fixed points are associated with the quasimomenta $k_m$ where $\bm{n}_{h}(k_m,t)$ points to the north (green) or south (magenta) pole of the Bloch sphere. These $k_m$ correspond to the intersections of the trajectory with the $z$ axis in Fig.~\ref{fig:fig2}.
  (b) In the sector of the critical momentum $k_c$, $\bm{n}_{h}(k_c,t)$ lies on the equator.
   After half a period of rotation, the spin vector is bound to be opposite in direction, where the time-evolved state becomes orthogonal to the initial state. These $k_c$ correspond to the intersections of the trajectory with the $x$ axis in Fig.~\ref{fig:fig2}.
  }
  \label{fig:fig3}
\end{figure}

\begin{figure*}[tbp]
  \centering
  \includegraphics[width=18cm]{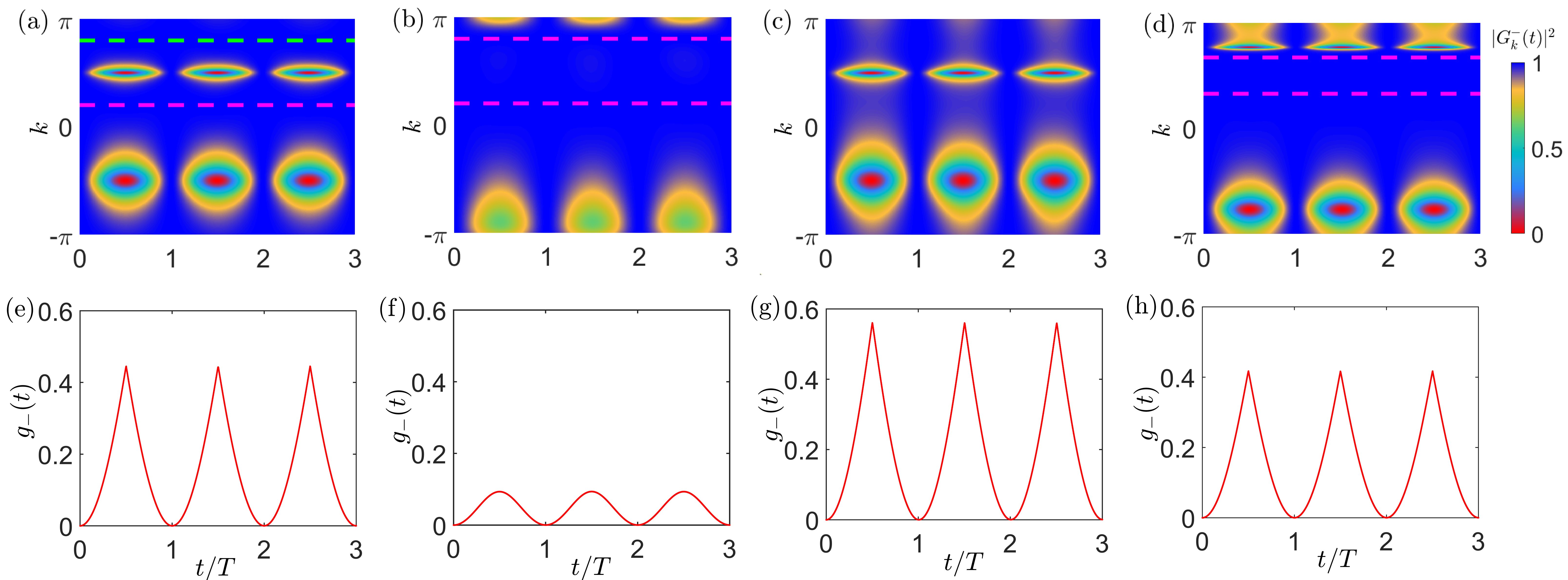}
  \caption{Loschmidt amplitude $|G^{-}_{k}(t)|^2$ and rate functions $g_-(t)$. (a)--(d) The density plot of $|G^{-}_{k}(t)|^2$ versus time $t$ and dimensionless quasimomentum $k$, where the parameters are the same as those in Figs.~\ref{fig:fig2}(a)--(d), respectively. The corresponding rate functions are shown in (e)--(h), and the $x$-axis labels for (a)--(d) are the same as (e)--(h). Floquet DQPTs occur in (a), (c), (d), where there are two critical quasimomentas in each period, and the corresponding rate functions appear non-analytic at the critical time. The magenta and green dashed horizontal lines corresponding to the locations of the two different types of fixed points.}
  \label{fig:fig4}
\end{figure*}

\subsection{Floquet dynamic quantum phase transitions}

We first examine the Floquet DQPT, its topological characterization, and its relation with the winding number $\nu$.

The DQPT is introduced as a temporal analog of continuous phase transitions. A DQPT occurs when the Loschmidt amplitude, defined as $G(t)=\langle\Psi(0)|\Psi(t)\rangle$, vanishes at critical times $t_c$. While the DQPTs are typically studied in the context of quench dynamics, they also emerge in Floquet dynamics~\cite{zhouFDQPT,FDQPT1,FDQPT2,FDQPT3}, and manifest as the  vanishing of $G(t)$ during the micromotion within each modulation period.

More specifically, assuming an initial state $|\psi^{-}(t)\rangle = \prod_{k}|\psi^{-}_{k}(t)\rangle $, the Loschmidt amplitude is
\begin{align}
G^{-}(t) = \langle\psi^{-}(0)|\psi^{-}(t)\rangle = \prod_{k}G^{-}_{k}(t),
\end{align}
with
\begin{align}
G^{-}_{k}(t) &= \langle\psi^{-}_{k}(0)|\psi^{-}_{k}(t)\rangle = e^{-i\epsilon^{-}_{k} t}\langle\mu^{-}_{k}|U^{k}_{F}(t)|\mu^{-}_{k}\rangle \nonumber\\
&=  e^{-i\epsilon^{-}_{k} t}[\frac{(|\epsilon_k|-h_{z}(k))^2 + h^{2}_{x}(k)e^{i\omega t}}{h^{2}_{x}(k) + (|\epsilon_k|-h_{z}(k))^2} \nonumber\\
&- \frac{h^{2}_{x}(k)(|\epsilon_k|-h_{z}(k))\frac{t''_{so}}{4\omega}(e^{2i\omega t}-e^{-i\omega t})}{h^{2}_{x}(k) + (|\epsilon_k|-h_{z}(k))^2}]. \label{eq:Loshmidtamplitude}
\end{align}

From the analytic expression (\ref{eq:Loshmidtamplitude}), we see that $G^-_k(t)$ associated with critical quasimomentum $k_c$ vanishes periodically at $t_c= (n+\frac{1}{2})T$, where $n=0,1,2,\ldots$. Here the critical quasimomentum $k_c$ satisfies the relation
\begin{align}
h_{z}(k_c) = h_{x}(k_c)\frac{[(t'_{so}-t_{so})\sin(k_c)-t''_{so}/2]}{2\omega}. \label{eq:Critialmomentum}
\end{align}
It follows that the rate function, the temporal analog of the free energy, $g_{-}(t)=-\frac{1}{2\pi}\int^{\pi}_{-\pi}dk ln|G^{-}_{k}(t)|^2$ becomes nonanalytic at critical times, which is the signature of the DQPT.

For a more transparent understanding of the Floquet DQPT and its connection with the winding number $\nu$, let us first consider the case where the first-order high-frequency corrections in $H_{\text{eff}}$ and $U^k_F(t)$ are negligible. In this case, $U^k_F(t)$ is diagonal, and, according to Eq.~(\ref{eq:F-dynamics}), the state dynamics in each $k$ sector can be simply understood as a rotation around the $z$ axis on the Bloch sphere, as indicated in Fig.~\ref{fig:fig3}. If the initial state lies along the $+z$ or $-z$ axis, the time-evolved state only collects a time-dependent phase. We therefore define these as two different types of fixed points. In Fig.~\ref{fig:fig2}, these fixed points correspond to the intersection of the elliptic trajectory with the $z$ axis, which takes place for $k_m=\{\arcsin[t''_{so}/2(t_{so}+t'_{so})],\pi-\arcsin[t''_{so}/2(t_{so}+t'_{so})]\}$ in our case.
On the other hand, at the critical momentum $k_c=\{\arccos[(m_z-\omega/2)/2t_s],2\pi-\arccos[(m_z-\omega/2)/2t_s]\}$, $h_z(k_c)= 0$ in the high-frequency limit [see Eq.~(\ref{eq:Critialmomentum})], corresponding to the intersection of the elliptic trajectory with the $x$ axis in Fig.~\ref{fig:fig2}. At these quasimomenta, the initial state lies on the equator and evolves to its orthogonal state on the opposite side of the equator at $t_c$ [see Fig.~\ref{fig:fig3}(b)].

By considering different configurations of the trajectory with respect to the $x$--$z$ axes (see Fig.~\ref{fig:fig2}), several conclusions can be drawn. First, if two distinct types of fixed points exist, as in Fig.~\ref{fig:fig2}(a), Floquet DQPTs necessarily occur. Since the critical quasimomenta correspond to the intersections of the elliptic trajectory with the $x$ axis, they necessarily exist when the trajectory simultaneously intersects with the $z$ axis on both the positive and negative sides. Second, if two distinct types of fixed points exist, the winding of the Floquet effective Hamiltonian is non-zero, since the origin is then enclosed by the trajectory. Note that if the dynamics features no fixed points [see Fig.~\ref{fig:fig2}(c)] or a pair of fixed points of the same type [see Fig.~\ref{fig:fig2}(b),~\ref{fig:fig2}(d)], DQPT may or may not occur, and there ceases to be a direct connection between the two. However, in this case, the topology of $H_{\text{eff}}$ is necessarily trivial.

In Fig.~\ref{fig:fig4}, we confirm the analysis above by explicit numerical simulations of the Floquet dynamics. Figures~\ref{fig:fig4}(a)--~\ref{fig:fig4}(d) correspond to the trajectories in Figs.~\ref{fig:fig2}(a)--~\ref{fig:fig2}(d), respectively. Correspondingly, the rate function $g_{-}(t)$ shows periodic nonanalyticities in Figs.~\ref{fig:fig4}(e),~\ref{fig:fig4}(g),~\ref{fig:fig4}(h). The effective Hamiltonian is topological only in the case of Fig.~\ref{fig:fig4}(a). Further, while both Floquet DQPTs and fixed points exist in Figs.~\ref{fig:fig4}(d) and ~\ref{fig:fig4}(h), they are not directly correlated.

\begin{figure*}[tbp]
  \centering
  \includegraphics[width=18cm]{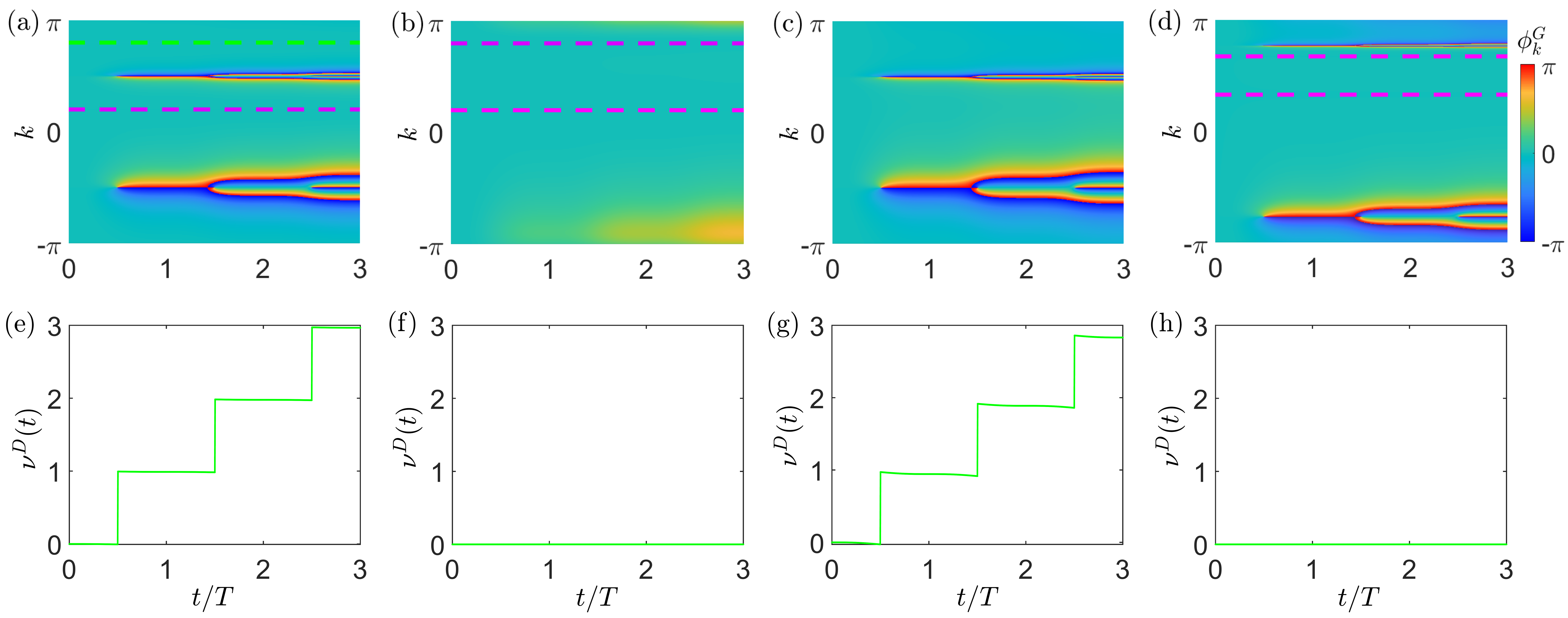}
  \caption{Pancharatnam geometric phases and DTOPs. (a)--(d) The density plot of the Pancharatnam geometric phases versus time $t$ and dimensionless quasimomentum $k$, where the parameters are the same as those in Figs.~\ref{fig:fig2}(a)--(d), respectively. The corresponding DTOPs are shown in (e)--(h), and the $x$-axis labels for (a)--(d) are the same as (e)--(h). The magenta and green dashed horizontal lines corresponding to the locations of the two different types of fixed points.}
  \label{fig:fig5}
\end{figure*}

Floquet DQPTs can be regarded as a dynamic topological phenomena, as they can be characterized by a dynamic topological order parameter, defined as~\cite{Heyldynamic}
\begin{align}
\nu^{D}(t)=\frac{1}{2\pi}\int^{k_{n}}_{k_{m}}dk\frac{\partial\phi^{G}_{k}(t)}{\partial k}, \label{eq:DTOP}
\end{align}
where $\phi^{G}_{k}(t)=\phi_k (t) - \phi^{\text{dyn}}_{k}(t)$ is the Pancharatnam geometric phase. Here the full phase factor $\phi_k$ is derived from the Loschmidt amplitude $G^{-}_{k}(t)=|G^{-}_{k}(t)|e^{i\phi_{k}(t)}$, and $\phi^{\text{dyn}}_{k}(t)=-\int^{t}_{0}ds\langle\psi^{-}_{k}(s)|H(k,s)|\psi^{-}_{k}(s)\rangle$ is the dynamic phase.
The integral is performed between two neighboring fixed points $k_m$ and $k_n$($m,n=1,2,\ldots $), where the Pancharatnam geometric phase vanishes at all times.
While $\nu^{D}(t)$ should be quantized under the existence of fixed points, it can change its value only at times when a critical quasimomentum $k_c$ exists in between $k_m$ and $k_{n}$, such that $G^{-}_{k_c}(t)=0$. This is satisfied when distinct fixed points exist, as is the case with Fig.~\ref{fig:fig3}(a) and Fig.~\ref{fig:fig4}(a).

In Fig.~\ref{fig:fig5}, we show the numerically calculated $\phi^{G}_{k}(t)$ and $\nu^D$ for the four cases in Figs.~\ref{fig:fig2} and~\ref{fig:fig4}. The abrupt jumps in the dynamic topological order parameter $\nu^D$ correctly reflects the onset of Floquet DQPTs in Fig.~\ref{fig:fig5}(e), while in Figs.~\ref{fig:fig5}(f),~\ref{fig:fig5}(h), $\nu^D$ remains zero as the fixed points are of the same type [see Figs.~\ref{fig:fig2}(b),~\ref{fig:fig2}(d)]. Further, $\nu^D$ is not quantized in Fig.~\ref{fig:fig5}(g), due to the absence of fixed points here [see Fig.~\ref{fig:fig2}(c)].

From the above discussions, we see that in the absence of the Floquet topology ($\nu=0$), there is no direct correlation between Floquet DQPTs, fixed points, and dynamic topological order parameter. Whereas the three are closely connected when $\nu\neq 0$. The emergence of dynamic topological phenomena are therefore protected by the Floquet topology.

\begin{figure*}[tbp]
  \centering
  \includegraphics[width=13.5cm]{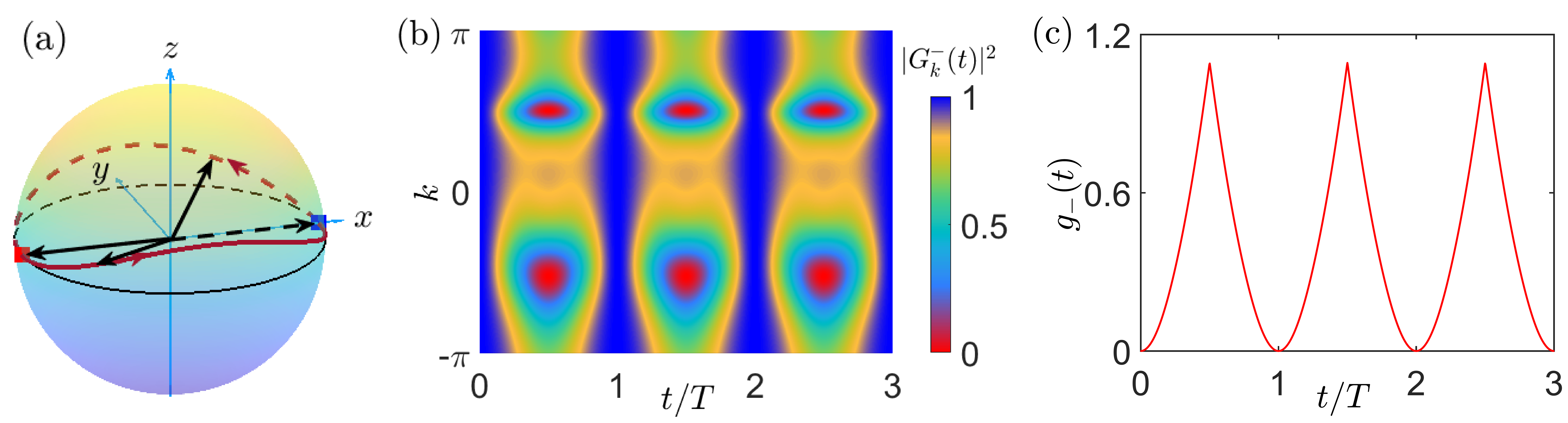}
  \caption{(a) Away from the high-frequency limit, the trajectory (dark red line) of $\bm{n}_{h}(k_c,t)$ at the critical quasimomentum $k_c$ forms a closed loop on the Bloch sphere, but no longer in the plane of the equator.
 The initial state and the state at the critical time are still orthogonal with each other. (b) and (c) are $|G^{-}_{k}(t)|^2$ and the rate function under the same parameters. The parameters are $m_z=5t_s$, $t_{so}=0.1t_s$, $t'_{so}=0.4t_s$, $t''_{so}=2.4t_s$, and $\omega = 10t_s$. For our calculations, we take $V_0 = 8E_R$, so that $t_s\approx 0.0375E_R$}.
  \label{fig:fig6}
\end{figure*}

\subsection{Away from the high-frequency limit}
\begin{figure*}[tbp]
  \centering
  \includegraphics[width=13.5cm]{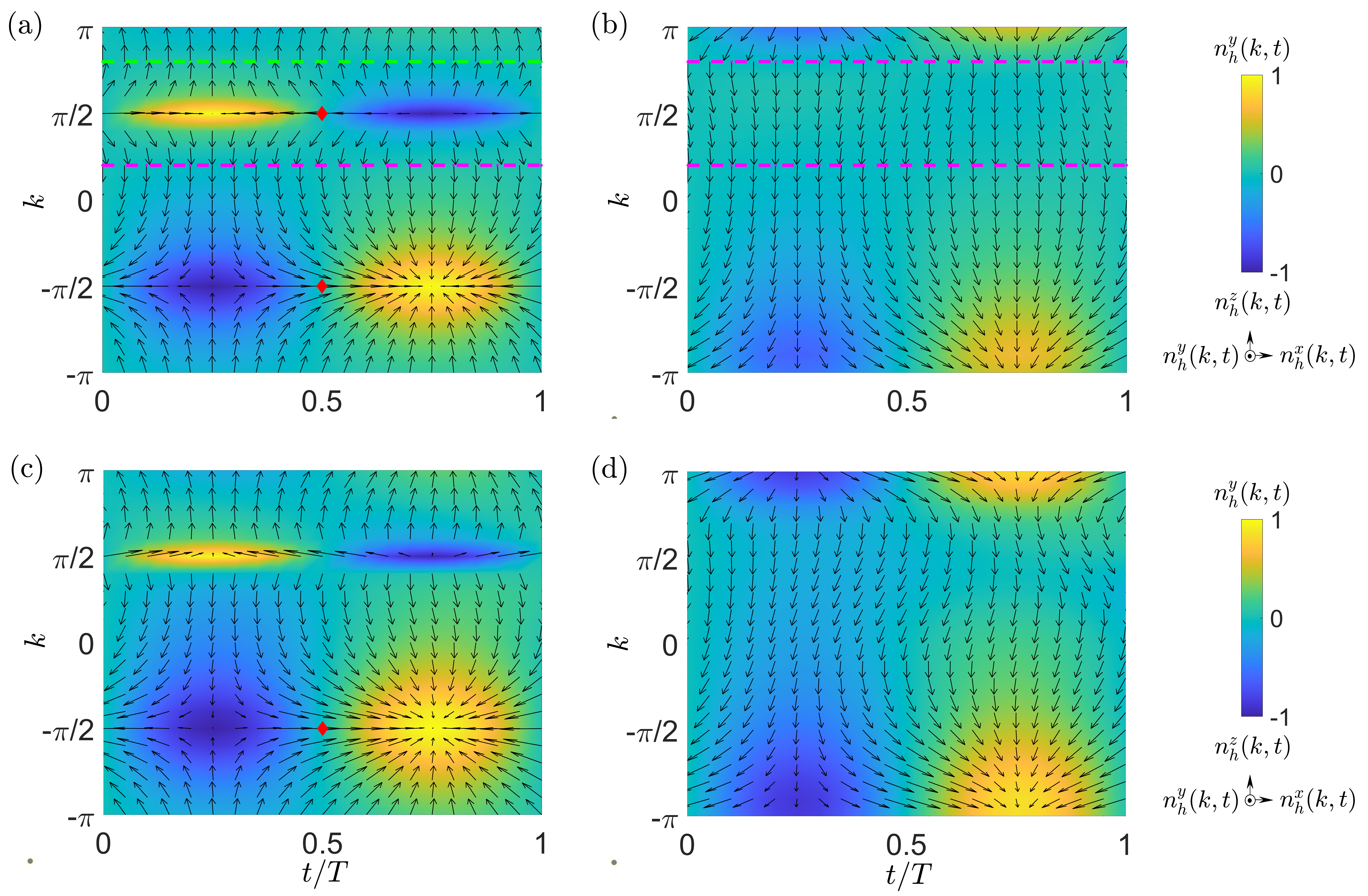}
  \caption{Spin texture $\bm{n}_h(k,t)$ in the $k$--$t$ domain. (a) The appearance of skyrmion lattices corresponding to the quantized dynamical Chern number. The parameters are the same as in those in Fig.~\ref{fig:fig2}(a). The magenta and green dashed horizontal lines indicate two different types of fixed points, respectively. (b) Spin texture under the parameters of Fig.~\ref{fig:fig2}(b), the two magenta dashed horizontal lines indicate two fixed points of the same type. (c) Spin texture under the parameters $m_z=5t_s$, $t_{so}=0.5t_s$, $t'_{so}=0.2t_s$, and $t''_{so}=1.2t_s$, where Floquet DQPTs exist but there are not fixed points. (d) Spin texture under the parameters $m_z=2.5t_s$, $t_{so}=0.5t_s$, $t'_{so}=0.2t_s$, and $t''_{so}=1.2t_s$, where neither fixed points nor Floquet DQPTs exist. In (a),(b), the $x$-axis labels are the same as (c),(d). In (c),(d), we take $V_0 = 8E_R$, $t_s\approx 0.0375E_R$, and $\omega = 10t_s$. Locations of DQPTs are marked by red dots in (a),(c), whereas there are no DQPTs in (b),(d).}
  \label{fig:fig7}
\end{figure*}

The analyses above are applicable in the high-frequency limit, where the first-order corrections are dropped in $U^k_F(t)$ and $H_{\text{eff}}$. The micromotion thus corresponds to a rotation around the $z$ axis on the Bloch sphere. When the first-order corrections are included, according to Eq.~(\ref{eq:micromotion}), $U_F^k(t)$ is no longer diagonal, and the micromotion corresponds to a general closed loop around the Bloch sphere, as indicated in Fig.~\ref{fig:fig6}(a). It follows that fixed points no longer exist, but Floquet DQPTs can still occur.
For instance, with the first-order corrections, when Eq.~(\ref{eq:Critialmomentum}) is satisfied at a critical momentum $k_c$, Floquet DQPTs occur at $t_c$, as the time-evolved state becomes orthogonal to the initial state in the corresponding $k_c$ sector. Such a situation is illustrated in Fig.~\ref{fig:fig6}.
Our result here complements previous studies of Floquet DQPTs under different setups~\cite{zhouFDQPT,FDQPT1,FDQPT2,FDQPT3}, where
the exact Floquet Hamiltonian can be derived without invoking the high-frequency expansion.
Beyond the first-order corrections, we have numerically confirmed that Floquet DQPTs generally exist, but the conditions for their occurrence needs further study.



\subsection{Dynamic skyrmions in Floquet System}

A related dynamic topological construction is the dynamic skyrmions, protected by the dynamic Chern number defined on the emergent momentum-time manifold.
In previous studies of quench dynamics~\cite{skyrmions2,skyrmions1,schen}, it is shown that when the initial and final Hamiltonians possess different topological invariants, fixed points of different types emerge in the dynamics, which divides the momentum-time manifold into various submanifolds on which dynamic skyrmion spin textures exist.

Similarly, for the Floquet dynamics in the high-frequency limit, a non-trivial topology of the Floquet Hamiltonian protects fixed points of different kinds. Based on these fixed points,
a dynamic Chern number can be defined~\cite{schen}
\begin{align}
C_{\text{dyn}}^{mn}=\frac{1}{4\pi}\int^{k_{n}}_{k_{m}}dk\int^{T}_{0}dt[\bm{n}_{h}(k,t)\times\partial_{t}\bm{n}_{h}(k,t)]\cdot\partial_{k}\bm{n}_{h}(k,t), \label{eq:DyChern}
\end{align}
where $\bm{n}_{h}(k,t)=\text{Tr}(|\psi^{-}_{k}(t)\rangle\langle\psi^{-}_{k}(t)|\bm{\sigma})$ is the normalized vector on the Bloch sphere corresponding to the stationary Floquet state with momentum $k$ at time $t$.
While $C_{\text{dyn}}^{mn}$ is always quantized in the presence of fixed points, it is nonzero when the neighboring fixed points $k_m$ and $k_{n}$ are of different types; and it vanishes when they are of the same type~\cite{schen,skyrmions1}. Since the dynamic Chern number is essentially the skyrmion number for the spin texture $\bm{n}_{h}(k,t)$, the topology of the Floquet Hamiltonian is directly related to Floquet DQPTs and dynamic skyrmions in the micromotion.
In Figs.~\ref{fig:fig7}(a),~\ref{fig:fig7}(b), we show the spin textures in the momentum-time domain within one modulation period under the parameters of Figs.~\ref{fig:fig2}(a),~\ref{fig:fig2}(b), respectively. Dynamic skyrmions only exist in Fig.~\ref{fig:fig7}(a), where both types of fixed points exist.

For the general Floquet dynamics away from the high-frequency limit, fixed points no longer exist in general. The dynamic Chern number is ill defined (not quantized), and dynamic skyrmion structures no longer exist. As concrete examples, in Figs.~\ref{fig:fig7}(c),~\ref{fig:fig7}(d), we illustrate two cases where either the Floquet DQPTs exist in the absence of fixed points [ Fig.~\ref{fig:fig7}(c)]; or neither of them exist [Fig.~\ref{fig:fig7}(d)].
%
%

\begin{figure}[tbp]
  \centering
  \includegraphics[width=7cm]{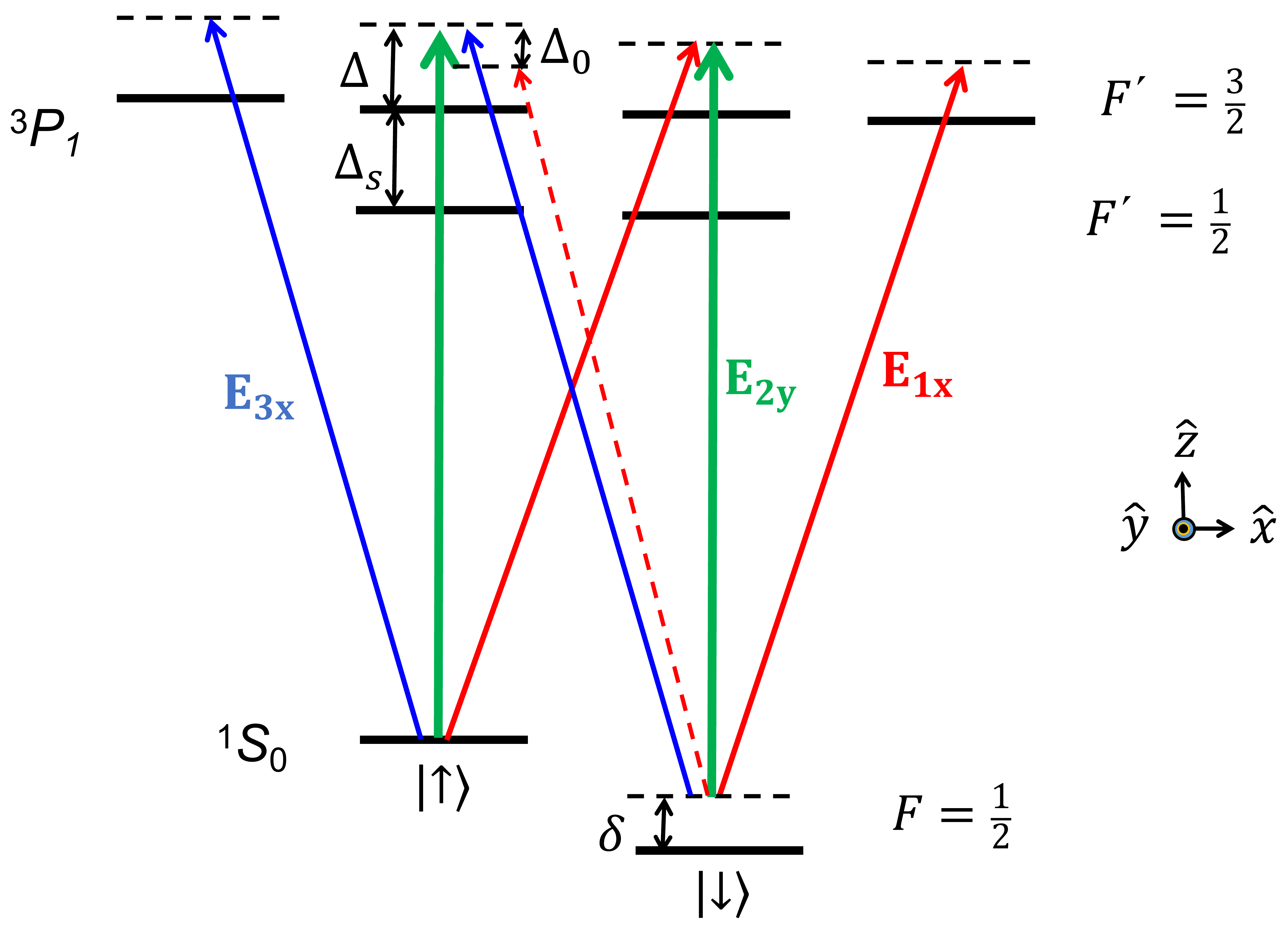}
  \caption{Schematic illustration of the coupling scheme for realizing the time period optical Raman lattice in alkaline-earth-like atoms $^{171}$Yb. The standing wave $\mathbf{E_{0x}}$ used to generate the 1D optical lattice potential in the $x$ direction is not shown in the figure. $F$ and $F'$ indicate different quantum numbers of hyperfine ground states and excited states, respectively, $\Delta_s$ is the hyperfine manifolds, and the $z$ axis is the quantization axis. }
  \label{fig:fig8}
\end{figure}

\section{Implementation}

Hamiltonian (\ref{eq:RamanH}) can be implemented using alkaline-earth(-like) atoms in an optical Raman lattice. As a concrete example, we consider the $\{^1S_0, ^3P_1\}$ manifolds of $^{171}$Yb atoms, and denote the hyperfine ground states as the two spins in (\ref{eq:RamanH}): $|\uparrow\rangle = |{}^1S_0,m_F=\frac{1}{2}\rangle$ and $|\downarrow\rangle = |{}^1S_0,m_F=-\frac{1}{2}\rangle$, as illustrated in Fig.~\ref{fig:fig8}.
A linearly polarized standing-wave laser with the electric-field vector $\mathbf{E_{0x}} = \hat{z}|E_{0x}|\cos(\pi k_0 x) e^{-i\omega_0 t}$ generates a one-dimensional optical lattice potential along the $x$ direction. The laser is blue detuned, with a detuning of $\sim 1$ GHz from the  $F=\frac{1}{2}\longleftrightarrow F'=\frac{3}{2}$ transition. The resulting lattice depth is given by
 $V_0 = \sum_{F'}|\Omega_{0x}^{\uparrow,F'}|^2/\Delta_{F'}=\sum_{F'}|\Omega_{0x}^{\downarrow,F'}|^2/\Delta_{F'}$.
Here the Rabi frequencies $\Omega_{0x}^{\sigma,F'}$ ($\sigma=\uparrow,\downarrow$) and the single-photon detunings $\Delta_{F'}$ are associated with the corresponding transitions between the hyperfine ground states and all relevant excited states with $F'=\frac{1}{2},\frac{3}{2}$ in the $^3P_1$ manifold.


The Raman potential $M_1(x,t)$ is generated by a Raman process consisting of a blue-detuned (with a detuning $\sim 2$ GHz) standing-wave laser $\mathbf{E_{1x}} = \hat{y}|E_{1x}|\sin(\omega t)\cos(\pi k_0 x) e^{-i\omega_1 t}$ with frequency $\omega_1$, and a plane-wave laser $\mathbf{E_{2y}} = \hat{z}|E_{2y}| e^{i(\pi k_0 y-\omega_2 t)}$ propagating along the $-y$ direction with frequency $\omega_2$. While both lasers are linearly polarized, they propagate along different directions, and hence drive different transitions, as illustrated in Fig.~\ref{fig:fig8}. The two lasers thus couple the two hyperfine ground states in a spatially dependent Raman process, with $M_1(x,t) = M_1\sin(\omega t)\cos(\pi k_0 x) e^{-i\pi k_0 y}$ and
$M_1 = \sum_{F'}\Omega_{1x}^{\uparrow,F'} \Omega_{2y}^{\downarrow,F'}/\Delta'_{F'}$, where
$\Delta'_{F'}$ is the single-photon detuning of the Raman coupling with respect to the $F'$ manifold, while $\Omega_{1x}^{\sigma,F'}$ and $\Omega_{2y}^{\sigma,F'}$ are the Rabi frequencies of the corresponding light field coupling the spin state $|\sigma\rangle$ in the ground state to the $F'$ manifold.
The effective Zeeman field $\delta$ in (\ref{eq:RamanH}) corresponds to the two-photon detuning.

The Raman potential $M_2(x,t)$ is generated by another Raman process consisting of $\mathbf{E_{2y}}$, and a circularly polarized plane-wave laser $\mathbf{E_{3x}} = (\hat{z}-i\hat{y})i|E_{3x}|\cos(\omega t) e^{-i(\pi k_0 x+\omega_3 t)}$ of frequency $\omega_3$. Together, they give $M_2(x,t) = M_2\cos(\omega t) [\cos(\pi k_0 x)+i\sin(\pi k_0 x)]e^{i\pi k_0 y}$, with $M_2 = \sum_{F'}\Omega_{2y}^{\uparrow,F'} \Omega_{3x}^{\downarrow,F'}/\Delta'_{F'}$. Here $\Omega_{3x}^{\sigma,F'}$ is the corresponding Rabi frequency of the circularly polarized laser coupling.
Note that the two-photon detuning $\delta$ is set to be the same as that of the first Raman process $M_1(x,t)$.

As is the case with recent experiments~\cite{Gyubongjo18,2Dsoc}, all Raman beams can be generated by  the same laser source, with their frequency difference $\delta\omega = \omega_2 - \omega_1 = \omega_3 - \omega_2$ tunable through an
acousto-optical modulator. We further consider imposing a magnetic field of $\sim 40$ G, such that the two-photon Raman coupling from $\mathbf{E_{1x}}$ (red dashed in Fig.~\ref{fig:fig1}) and $\mathbf{E_{2y}}$ is far-detuned, with a two-photon detuning $\delta_0=2\delta\omega+\delta\approx 16 E_R$. Note that the typical Raman coupling strength does not exceed $2E_R$, with the recoil energy $E_R = \pi^2 k^2_0/2m \approx 3.77$kHz for $^{171}$Yb atoms. Further, when the amplitudes of the light fields $|E_{0x}|$ and $|E_{2y}|$ are significantly stronger than $|E_{1x}|$ and $|E_{3x}|$, the optical-dipole potential generated by the latter can be neglected. We thus arrive at Hamiltonian (\ref{eq:RamanH}), apart from an irrelevant phase factor $e^{\pm i\pi k_0 y}$.

While the synthetic topology in the low-frequency limit can be experimentally detected similar to the recent observation of Thouless pumping in cold atoms~\cite{pumpexp1,pumpexp2}, the dynamic skyrmion structures can be probed using a combination of time-of-flight imaging and momentum-resolved interference~\cite{2ddqptatoms}, which would resolve the full-state evolution.

\section{Summary}
To summarize, we propose to implement a Raman lattice with periodic modulation using alkaline-earth(-like) atoms.
The system is capable of simulating topological phenomena at different ends of the modulation frequency. In the low-frequency limit, the adiabatic dynamics of the system can be mapped to a two-dimensional topological insulator. In the high-frequency limit, various dynamic topological phenomena emerge and are closely related through the topology of the Floquet Hamiltonian. The proposed system thus offers a versatile quantum simulation platform for synthetic topology. For future studies, it would be interesting to explore periodically driven lattice potentials in higher spatial dimensions, where the additional spatial degrees of freedom provides the room for engineering even richer dynamic topological structures. While Floquet DQPTs have been reported under general driving frequencies~\cite{zhouFDQPT,FDQPT1,FDQPT2,FDQPT3},
it would be interesting to explore whether a universal criterion exists for their existence.

\begin{acknowledgments}
We thank Wei Zheng for helpful comments. This work is supported by  the National Natural Science Foundation of China (11974331), and the National Key Research and Development Program
of China (2016YFA0301700, 2017YFA0304100).
\end{acknowledgments}

\appendix

\section{Floquet Hamtiltonian}

Following the Floquet theory, the time-evolved state under the time periodic tight-binding Hamiltonian in Eq.~(\ref{eq:HTI}) can be written as
\begin{align}
|\psi_{\alpha}(t)\rangle = e^{-i\epsilon_{\alpha}t}|u_{\alpha}(t)\rangle,
\end{align}
where the Floquet mode $|u_{\alpha}(t)\rangle=|u_{\alpha}(t+T)\rangle$ is periodic in time with a quasienergy $\epsilon_{\alpha}$. The time-dependent Schr\"{o}dinger's equation can be written as
\begin{align}
H_{\text{FT}}|u_{\alpha}(t)\rangle = \epsilon_{\alpha}|u_{\alpha}(t)\rangle,
\end{align}
where $H_{\text{FT}}=H_{\text{T}}-i\frac{\partial}{\partial t}$. We then perform a Fourier transform on the Floquet mode
\begin{align}
u_{\alpha}(t) = \sum_{n}c_{n}e^{in\omega t},
\end{align}
which is equivalent to writing $c_{j,\sigma}=e^{-in\omega t}c_{n,j,\sigma}$ in Eq.~(\ref{eq:HTI}).
The Floquet operator $H_{\text{FT}}$ is then cast into the form of Eq.~(\ref{eq:FloquetHTI}). Note that the last term of Eq.~(\ref{eq:FloquetHTI}) comes from the term $-i\frac{\partial}{\partial t}$.

\section{Effective Floquet Hamiltonian and micromotion}

In order to get the effective Floquet Hamiltonian $H_{\text{eff}}$, we perform a Fourier transform on the spatial modes of (\ref{eq:HTI}), to get the time-dependent Bloch Hamiltonian
\begin{align}
H(k,t) = & \cos(\omega t)[2t'_{so}\sin(k) - t''_{so}]\sigma_x + 2t_{so}\sin(\omega t)\sin(k) \nonumber\\
&\cdot\sigma_y + [m_z - 2t_s\cos(k)]\sigma_z, \label{eq:BlochHkt}
\end{align}
where $c_{k,\sigma} = \frac{1}{\sqrt{N_{x}}}\sum_{j}e^{-ijk}c_{j,\sigma}$, with the dimensionless quasimomentum $k \in [-\pi,\pi)$ in units of $k_0$.

We then take a gauge transformation $U_{R}(t)=e^{i\omega (\sigma_0 -\sigma_z)t/2}$ on Eq.~(\ref{eq:BlochHkt}), and get the Bloch Hamiltonian in the corresponding rotating frame
\begin{align}
 H_{R}(k,t) &\equiv U_{R}^{\dag}(t)H(k,t)U_{R}(t) - iU_{R}^{\dag}(t)\frac{d}{dt}U_{R}(t)\nonumber\\
 &= h_{x}(k)\sigma_x  + h_{+}(k)e^{2i\omega t}\sigma_{+} + h_{-}(k)e^{-2i\omega t}\sigma_{-} \nonumber\\
 &+ h_{z}(k)\sigma_{z} + \frac{\omega}{2}\sigma_{0}. \label{eq:BlochHR}
\end{align}
Here $\sigma_{\pm}=(\sigma_x \pm \sigma_y)/2$ are ladder operators, and $h_{+}(k)=h_{-}(k)=(t'_{so}-t_{so})\sin(k)-t''_{so}/2$.
The frequency components of $H_{R}(k,t)$
are
\begin{align}
 H_{m} = \frac{1}{T}\int_{0}^{T}dt e^{-im\omega t}H_{R}(k,t).
\end{align}
The only non-vanishing components are: $H_0 = [m_z - \omega/2 - 2t_s\cos(k)]\sigma_{z} + [(t'_{so}+t_{so})\sin(k)-t''_{so}/2]\sigma_x$, $H_2=h_{+}(k)\sigma_{+}$ and $H_{-2}=h_{-}(k)\sigma_{-}$.
The frequency components are useful for evaluating Eqs.~(\ref{eq:Heff}) and (\ref{eq:micromotion}).

\end{document}